# *Parallel, Series, and Intermediate Interconnections of Optical Nanocircuit Elements*

# *Part 1: Analytical Solution*

*Alessandro Salandrino, Andrea Alù, Nader Engheta*

University of Pennsylvania

Department of Electrical and Systems Engineering

200 South 33rd Street, Philadelphia, Pennsylvania 19104, USA

## Abstract

Following our recent development of the paradigm for extending the classic concepts of circuit elements to the infrared and optical frequencies [N. Engheta, A. Salandrino, A. Alù, *Phys. Rev. Lett.* **95**, 095504 (2005)], in this paper we investigate the possibility of connecting nanoparticles in series and in parallel configurations, acting as nanocircuit elements, In particular, we analyze a pair of conjoined half-cylinders, whose relatively simple geometry may be studied and analyzed analytically. In this first part of the work, we derive a closed-form quasi-static analytical solution of the boundary-value problem associated with this geometry, which will be applied in Part II for a nanocircuit and physical interpretation of these results.

## *1. Introduction*

The advances in the science and engineering of nanofabrication and characterization techniques over the past few decades have provided the possibility for exploring optical and electronic phenomena at the nanometer scales with precision previously unimagined [1]. Among the optical phenomena characterizing the nano scale domain, surface plasmons have attracted a great deal of attention in the science, applied science and engineering communities.

Plasmonic materials indeed offer interesting possibilities in manipulation of optical fields, ranging from scattering enhancement, to sub-diffraction confinement and guidance. These effects are indeed striking and are not ordinarily observed with normal dielectrics, so they are often called "anomalous", even though they may be predicted and described within the framework of the classical scattering theories [2]-[3] and solid state physics [4]. It is important to point out that in this case the fabrication techniques are leading ahead of the design techniques, and therefore new design ideas are needed in order to fully exploit many possibilities provided by the current technologies. Many groups all over the world have concentrated their efforts in the area of nanotechnology [5]-[15], with an ever increasing number of suggestions and ideas for applications and devices based on control of plasmonic resonances.

We have recently conducted studies for developing and extending the concept of circuit to the optical domain [16]. That implied a partial redefinition of the relevant electrical quantities commonly involved in the low frequency designs, shifting the focus from the conduction currents to the displacement currents flowing in an optical nanocircuit. The operation through displacement currents rather than conduction currents has major consequences in the way the circuit elements may be connected. As a matter of fact, the nature of the displacement currents makes it more difficult to confine them through specific paths and directions, as it is done by the physical boundary of a conductor when an actual flow of charge carriers is concerned. To this end, an important extension of our nanocircuit theory may reside in the possibility of interconnecting multiple nanocircuit elements together to form a complex nanocircuit system. The focus of the present work is on the first step towards such a goal, i.e., the analysis of the series and parallel interconnections between two nanocircuit elements, in the sense defined in the framework of our optical nanocircuit theory [16]. As we describe in detail in the second part of this paper [17], in which we focus on the physical aspects behind this interconnection and its role in the nanocircuit framework, one relevant geometry in this sense consists of a pair of two conjoined half-cylinders, which, depending on the orientation of the impinging electric field, may act as a parallel or a series combination of nanocircuit elements.

In this first part, we present a detailed electromagnetic solution for the quasi-static scattering problem associated with this geometry. In particular, here we derive a novel closed-form solution for this specific geometry, which has a simple physical interpretation in terms of the resonances associated with the coupling of the two nanocircuit elements. Moreover, the field distributions obtained through this analysis fully confirm the nanocircuit interpretation of the geometry under analysis. The results of this first part will therefore be of extreme use in the second part of this work [17], in showing how such a simple geometry may act as a basic interconnection between two nanocircuit elements, as an example for a more complex nanocircuit system.

Throughout the following analysis an $e^{-i\omega t}$ time dependence is assumed.

## 2. *Geometry of the Problem: a Dielectric Conjoined Cylinder*

The geometry we consider in the following is shown in Figure 1, and it consists of two half circular cylinders of radius $R$, cut perpendicularly to their circular cross-section along a diameter and conjoined along the cut. Under the assumption that the size of the structure is sufficiently smaller than the operating wavelength, which is a valid assumption for the typical size of the nanocircuit elements of interest in this work, the following analysis may be performed in the quasi-static approximation. It may be underlined that this assumption does not affect the necessary time variation of the involved phenomena, required for the definition of displacement current and for having plasmonic materials [18], but it rather limits the spatial variation of the fields across each nanoparticle, due to its small electrical size.

The structure is excited by a uniform electric field of amplitude $\mathbf{E}_0$ incident at an angle $\gamma$ with respect to the internal interface of the cylinder.

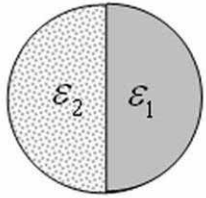

Figure 1 – Geometry of the problem: two conjoined half-cylinders of different permittivities

The goal of this part of our manuscript is to derive the distribution of the electric potential $\Phi(\rho,\varphi)$ inside and around the structure of Fig. 1. Such a potential is the solution of the Laplace equation in a suitable cylindrical reference system $(\rho,\varphi)$:

$$\rho^2 \frac{\partial^2 \Phi(\rho,\varphi)}{\partial \rho^2} + \rho \frac{\partial \Phi(\rho,\varphi)}{\partial \rho} + \frac{\partial^2 \Phi(\rho,\varphi)}{\partial \varphi^2} = 0. \tag{1}$$

Here we have used the translational symmetry of the 2-D problem at hand, for which the potential is uniform along the $z$ axis, i.e., the axis of the cylinder.

## 3. Formulation of the Dual Problem via Kelvin Transformation

The boundary value problem shown in Fig 1 cannot be easily handled in its original form because a set of orthogonal eigenfunctions conforming to such geometry is not readily available. For this reason our approach is to operate a conformal mapping of the original geometry into a new form which may be more easily solved.

In the following, we solve the boundary value problem of Fig. 1 by applying a conformal Kelvin transformation [19] :

$$\rho_K = \alpha^2/\rho, \qquad (2)$$

which maps the geometry of interest into a related dual problem. This transformation belongs to the inversions with respect to an analytic curve, and specifically consists of an inversion with respect to the circle of radius $\alpha$, called circle of inversion and centered at the origin of our cylindrical reference system. In general, the mapping between the original geometry and the transformed geometry is depicted in Fig. 2, where a circle (blue line, passing through points 1 and 2) is mapped into another circle (red line, passing through points $1_K$ and $2_K$) after applying (2).

The circle of inversion, represented by the dotted black line, and centered at the origin, maps any analytic curve inside its area into another analytic curve outside it. In particular, the mapping is invariant for circular shapes, as Fig. 2a shows, and a (blue) circle of radius $R$ centered at the Cartesian coordinates $\{0, r_0\}$ inside the (dotted) circle of inversion is mapped into a (red) circle of radius $R_K = \alpha^2 R/|r_0^2 - R^2|$ and center $\{0, \alpha^2 r_0/|r_0^2 - R^2|\}$ positioned outside of it. We note how in general the transformed circles are scaled in size with the quantity $\alpha^2/|r_0^2 - R^2|$ and reversed horizontally with respect to the original circles [the points 1 and 2 on the original (blue) circle are, respectively, mapped into the points $1_K$ and $2_K$ on the transformed (red) circle].

In the special case where the original (blue) circumference passes through the center of the circle of inversion, i.e., when $r_0 = R$, the mapped circle (red line) degenerates into the line $y = \alpha^2/(2R)$, as shown in Fig. 2b. This particular case allows us to map any circumference into a straight line, which may simplify the original boundary-value problem of Fig. 1 into the dielectric wedge problem of Fig. 3.

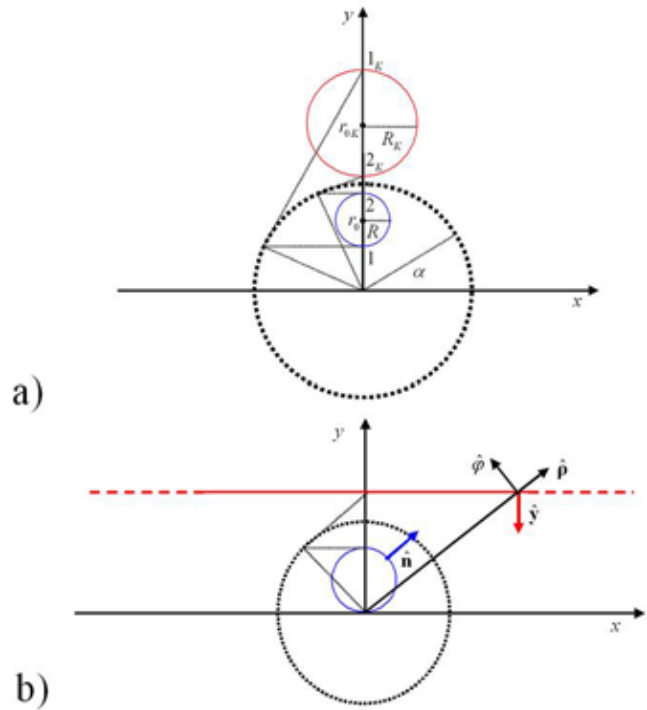

Figure 2 – (Color online). Schematic diagram of the mapping represented by Eq. (2). (a) The original blue circle, located inside the circle of inversion (dotted), is mapped into the red circle. (b) Degeneration of the mapped circle in the case in which the original circumference passes through the center of the circle of inversion.

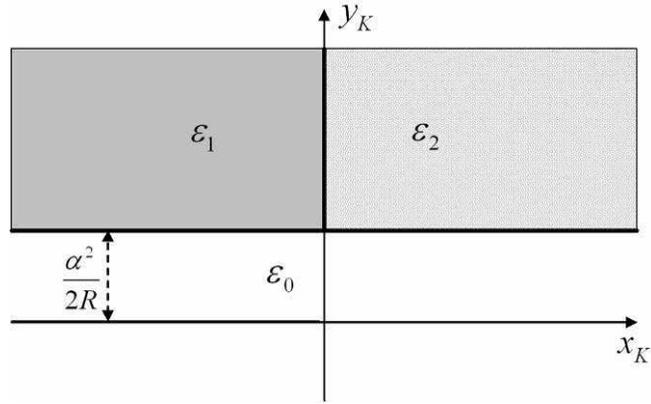

Figure 3 – Mapping of the geometry of Fig. 1, after applying the Kelvin transformation (2) in the specific case that is illustrated in Fig. 2b, in order to obtain the double-wedge rectangular problem.

One of the important properties of the Kelvin transformation is that it is invariant with Laplace's equation [19]:

$$\nabla^2 \Phi(\rho, \varphi) = 0 \Leftrightarrow \nabla^2 \Phi\left[\rho_K(\rho), \varphi\right] = 0. \tag{3}$$

This ensures that a solution of Laplace's equation in the mapped geometry is still a harmonic function in the original geometry.

The electromagnetic equivalence of the mapped problem is obtained by further showing that there is a one-to-one correspondence between the boundary conditions in the original domain and in the mapped domain. The Kelvin transformation, like most of the conformal mappings, is generally applied to problems involving only metallic boundaries or, in other words, Dirichlet boundary conditions where the consistency of the solutions in the two domains is guaranteed by the continuity of the mapping. Our geometry, on the other hand, involves interfaces between different dielectrics and, to the best of our knowledge, the Kelvin transform has not been applied in the past to this kind of problems.

Again, the continuity of the potential at any interface is guaranteed by the continuity of the mapping, but the continuity of the normal component of the electric displacement **D** at the interface between two different media is not as immediately evident. For this reason, in order to prove the complete correspondence of the original and the dual problem, in the next section we will prove the consistency of the boundary conditions between the two geometries.

## *4.    Electromagnetic Consistency of the Kelvin Transformation for the Present Problem*

The continuity of the potential is clearly maintained after the Kelvin transformation, since imposing that the solution is continuous along the three interfaces separating the three materials in Fig. 3 automatically implies that the conformal solution is continuous along the corresponding interfaces also in the original geometry of Fig. 1.

More intricate is the boundary condition on the normal component of the electric displacement vector **D**. First we note that the vertical line at the interface between the two half cylinders of Fig. 1 does not change its orientation after the mapping into Fig. 3 and that the Kelvin transformation involves only the radial variable $\rho$. This implies that the normal to the interface is parallel to $\hat{\boldsymbol{\varphi}}$ in both geometries and that $\partial \Phi(\rho,\varphi)/\partial \varphi = \partial \Phi\left[\rho_K(\rho),\varphi\right]/\partial \varphi$. The boundary condition at this interface therefore remains satisfied after the mapping.

Fig. 2b shows the orientation of the unit vector $\hat{\mathbf{n}}$ normal to the original circumference and of the transformed unit vector $\hat{\mathbf{y}}$. In particular along the circumference and along the line $y = \alpha^2/(2R)$ we get:

$$\begin{aligned}\hat{\mathbf{n}} &= \hat{\boldsymbol{\rho}} \sin\varphi - \hat{\boldsymbol{\varphi}} \cos\varphi \\ \hat{\mathbf{y}} &= \hat{\boldsymbol{\rho}} \sin\varphi + \hat{\boldsymbol{\varphi}} \cos\varphi\end{aligned}. \qquad (4)$$

We have to prove that

if: $\varepsilon_{in} \nabla \Phi_{in} \left[ \rho_K(\rho), \varphi \right] \cdot \hat{\mathbf{y}} = \varepsilon_0 \nabla \Phi_0 \left[ \rho_K(\rho), \varphi \right] \cdot \hat{\mathbf{y}}$ (5)

on the boundary $y = \alpha^2/(2R)$ in the mapped geometry, where $\varepsilon_{in}$ and $\Phi_{in}$ are, respectively, the permittivity and the potential in one of the two materials and $\Phi_0$ is the potential in the outer region, then: $\varepsilon_{in} \nabla \Phi_{in}(\rho, \varphi) \cdot \hat{\mathbf{n}} = \varepsilon_0 \nabla \Phi_0(\rho, \varphi) \cdot \hat{\mathbf{n}}$ (6) on the original (blue) circumference of Fig. 2b.

To this end, Eq. (5) can be rewritten as:

$$\varepsilon_{in} \left[ \frac{\partial \Phi_{in}[\rho_K(\rho),\varphi]}{\partial \rho} \sin\varphi + \frac{1}{\rho} \frac{\partial \Phi_{in}[\rho_K(\rho),\varphi]}{\partial \varphi} \cos\varphi \right] = \varepsilon_0 \left[ \frac{\partial \Phi_0[\rho_K(\rho),\varphi]}{\partial \rho} \sin\varphi + \frac{1}{\rho} \frac{\partial \Phi_0[\rho_K(\rho),\varphi]}{\partial \varphi} \cos\varphi \right]$$
. (7)

Applying transformation (2) we obtain:

$$\varepsilon_{in} \left[ \frac{\partial \Phi_{in}(\rho,\varphi)}{\partial \rho} \sin\varphi + \frac{1}{\rho} \frac{\partial \Phi_{in}(\rho,\varphi)}{\partial \varphi} \cos\varphi \right] = \varepsilon_0 \left[ \frac{\partial \Phi_0(\rho,\varphi)}{\partial \rho} \sin\varphi + \frac{1}{\rho} \frac{\partial \Phi_0(\rho,\varphi)}{\partial \varphi} \cos\varphi \right], \quad (8)$$

which coincides with (6), confirming that the boundary value problem in Fig. 3 is consistent with the one depicted in Fig. 1. This shows that it is sufficient to solve for the potential in one of the two geometries in order to determine the electromagnetic solutions in the other mapped geometry.

## 5.    *Transformation of the Incident Fields*

In order to complete the formulation of the dual problem the incident field also needs to be properly transformed according to the Kelvin mapping. If we assume that the conjoined half-cylinders are exposed to a uniform electric field linearly polarized along the line forming a general angle $\gamma$ with respect to the normal to the interface between the two half-cylinders, then the distribution of impressed (i.e., "incident") potential in the original geometry is given by $\Phi_{inc}(\rho,\varphi) = -E_0 \rho \cos(\varphi-\gamma)$. As we will show and discuss in detail in Part II, the conjoined half-cylinders of Fig. 1 may be seen as interconnected in series or in parallel as a function of the orientation of the impinging field, and in particular the two special cases of $\gamma = 0$ and $\gamma = \pi/2$ correspond to the series and the parallel connection.

After the Kelvin mapping, the impressed potential is transformed into:

$$\Phi_{inc}\left[\rho_K(\rho),\varphi\right] = -\frac{2\pi\varepsilon_0 E_0 \alpha^2}{2\pi\varepsilon_0} \frac{\cos(\varphi-\gamma)}{\rho_K}. \tag{9}$$

This expression coincides with the potential of a dipole orinted along the angle $\gamma$ with respect to the axis $x_K$, with dipole moment $p = 2\pi\varepsilon_0 E_0 \alpha^2$ and located at the origin [18]. Not surprisingly, this is physically related to the properties of the Kelvin transformation, which maps points placed at infinity in the original reference system into the origin of the conformal reference system. Since a uniform electric field may be considered to be generated by two equal and opposite charges located at two points at infinity, located symmetrically with respect to the origin, and the Kelvin transformation does not affect the angular distribution. This explains how these two charges collapse into a localized dipole at the origin in the conformal reference system, and oriented along the same angle with respect to the $x_K$ axis, as the original field makes with the $x$ axis (normal to the interface between the half-cylinders).

## 6. *Solution of the dual problem: the Double Dielectric Wedge*

The geometry and the excitation of the dual problem have been completely described in the previous section. We now proceed to its formal solution. First of all it is convenient to translate the double-wedge boundary shown in Figure 3 on the $x_K$ axis, with a new transformation rule given by:

$$\begin{cases} Y' = y_K - \dfrac{\alpha^2}{2R} \\ X' = x_K \end{cases} \qquad \begin{cases} \rho' = \sqrt{X'^2 + Y'^2} \\ \varphi' = \arctan\left(\dfrac{Y'}{X'}\right) \end{cases} \tag{10}$$

The boundary-value problem of Fig. 1 has been reduced to that of a double dielectric wedge excited by an electric dipole placed at the distance $r_0 = \alpha^2/(2R)$ from the interface of the double wedge with free space, as depicted in Figure 4. This problem may be solved for the translated potential $\Phi'(\rho', \varphi')$ under this quasi-static assumption, by applying the Mellin transformation, as suggested in [20].

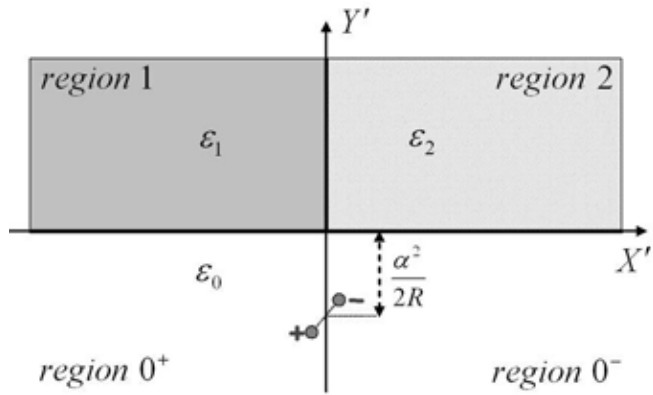

Figure 4 – Transformed boundary-value problem after the Kelvin transform (2) and the translation (10): a double-wedge excited by an electric dipole forming an angle $\gamma$ with the $X'$ axis equal to the angle formed by the uniform electric field with the normal to the interface between the two 2D half-cylinders in the original geometry (Fig. 1).

After having solved the rectangular problem of Fig. 4 for the potential $\Phi'(\rho', \varphi')$, we apply the inverse translation (10) and the inverse Kelvin transform (2) to return to the original problem and evaluate the potential distribution $\Phi(\rho, \varphi)$ for the conjoined half-cylinders of Fig. 1. In particular, this distribution assumes a closed-form expression when the permittivities of the two half-cylinders are equal and opposite in sign, which, as we discuss in the next section, corresponds to a resonant configuration.

Applying the Mellin's integral transform [20], defined as:

$$\tilde{f}(s) = \int_0^\infty (\rho')^{s-1} f(\rho') d\rho'$$
$$f(\rho') = \frac{1}{2\pi i} \int_{c-i\infty}^{c+i\infty} (\rho')^{-s} \tilde{f}(s) ds \quad , \quad (11)$$

we can expand the integral $\Phi'(\rho', \varphi')$ of the Laplace equation in the double wedge problem of Fig. 4 into a set of analytical functions $\tilde{\Phi}'(s, \varphi')$, depending on the Mellin radial frequency $s$ and on the angular coordinate $\varphi'$. In particular, in the four regions of Fig. 4 we get:

$$\begin{aligned}
\tilde{\Phi}'_{0^+}(s,\varphi') &= A_{0^+}(s)\cos(s\varphi') + B_{0^+}(s)\sin(s\varphi') \\
\tilde{\Phi}'_{0^-}(s,\varphi') &= A_{0^-}(s)\cos(s\varphi') + B_{0^-}(s)\sin(s\varphi') \\
\tilde{\Phi}'_{1}(s,\varphi') &= A_{1}(s)\cos(s\varphi') + B_{1}(s)\sin(s\varphi') \\
\tilde{\Phi}'_{2}(s,\varphi') &= A_{2}(s)\cos(s\varphi') + B_{2}(s)\sin(s\varphi')
\end{aligned} \quad . \quad (12)$$

The unknown coefficients $A$ and $B$ in each region are determined by imposing the boundary conditions at the three interfaces of Fig. 4. This allows the determination of the potential $\tilde{\Phi}'(s,\varphi')$ in the Mellin domain, that is related to the potential distribution $\Phi'(\rho',\varphi')$ in the double-wedge problem through (11), by applying an inverse Mellin transform with a residue evaluation of the integral [20]. The convergence of the integrals in (11) for the present configuration is guaranteed by the form of excitation, i.e., an electric dipole, whose radial dependence satisfies the convergence requirements [20].

It may be shown that the poles of the coefficients in (12), which determine the location of the residues in the general case, and therefore the distribution of the transformed potential, are given by the following dispersion equation in the general case:

$$\Delta(s) = \left[ 4\varepsilon_0\varepsilon_1\varepsilon_2 + \varepsilon_0^2(\varepsilon_1 + \varepsilon_2) + \varepsilon_1\varepsilon_2(\varepsilon_1 + \varepsilon_2) + (\varepsilon_0 + \varepsilon_1)(\varepsilon_0 + \varepsilon_2)(\varepsilon_1 + \varepsilon_2)\cos(\pi s) \right] \sin^2(\pi s/2) = 0. \qquad (13)$$

It must be noticed how in the special case of $\varepsilon_1 = \varepsilon_2 = \varepsilon_0$ the dispersion equation yields the set of solutions $s_N = N$, with $N$ being any integer, which corresponds to the poles of the original potential distribution, given by an electric dipole, consistent with the fact that in this case the conjoined half-cylinders and the corresponding double wedge are not present. When $\varepsilon_1 = -\varepsilon_2$, instead, independent of their relative value with respect to $\varepsilon_0$, Eq. (13) admits the set of solutions $s_N = 2N$, which corresponds to the potentials from two electric dipoles, one placed in the original source position, again corresponding to the impressed potential, the other placed and oriented symmetrically with respect to the origin, which is an image of the original source. As we discuss further in details in the second part of this manuscript, this special condition corresponds to a resonant configuration for the conjoined geometry under analysis.

## 7.   Inversion of the Mellin's Transform in the resonant case

In the resonant case $\varepsilon_2 = -\varepsilon_1$, the coefficients appearing in the formulas (12) have the following simple expressions:

$$A_{0^+} = A_1 = \frac{p\, r_0^{s-1} \sin(\gamma)}{2\varepsilon_0 \sin(s\pi/2)}$$

$$B_{0^+} = -\frac{p\, r_0^{s-1} \cos(\gamma)}{2\varepsilon_0 \sin(s\pi/2)}$$

$$A_{0^-} = -\frac{p}{\varepsilon_0} r_0^{s-1} \cos(\gamma)\cos(s\pi/2) + \frac{p\, r_0^{s-1} \sin(\gamma)\cos(s\pi)}{2\varepsilon_0 \sin(s\pi/2)}$$

$$B_{0^-} = -\frac{p\, r_0^{s-1} \cos(\gamma)\cos(s\pi)}{2\varepsilon_0 \sin(s\pi/2)} - \frac{p}{\varepsilon_0} r_0^{s-1} \sin(\gamma)\cos(s\pi/2)$$

$$B_1 = -\frac{p\, r_0^{s-1} \cos(\gamma)}{2\varepsilon_1 \sin(s\pi/2)}$$

$$A_2 = -\frac{p\, r_0^{s-1} \cos(\gamma)\cos(s\pi/2)}{\varepsilon_1} + \frac{p\, r_0^{s-1} \sin(\gamma)\cos(\pi s)}{2\varepsilon_0 \sin(s\pi/2)}$$

$$B_2 = \frac{p\, r_0^{s-1} \cos(\gamma)\cos(s\pi)}{2\varepsilon_1 \sin(s\pi/2)} + \frac{p}{\varepsilon_0} r_0^{s-1} \sin(\gamma)\cos(s\pi/2) \tag{14}$$

In the inversion of the Mellin transform, only the terms with poles give contributions to the potential through their residues. These poles, consistent with (13), are located at $s = 2N$ and the transformed potential distribution may be written in compact form as:

$$\tilde{\Phi}'_i(s,\varphi') = r_0^{s-1} \frac{p}{\varepsilon_0} \frac{\varepsilon_i \sin(\gamma)\cos(s\varphi') - \varepsilon_0 \cos(\gamma)\sin(s\varphi')}{2\varepsilon_i \sin(s\pi/2)}, \tag{15}$$

with $i = 0, 1, 2$ in the different regions of space.

The inversion of the Mellin transform may be performed by applying the residue theorem, once a proper Bromwich contour is chosen [20], resulting in:

$$\Phi'_i(\rho',\varphi') = \mathrm{sgn}(\rho' - r_0)\, p \sum_{N=0}^{\infty} \frac{\varepsilon_i \sin(\gamma)\cos(2N\varphi') - \varepsilon_0 \cos(\gamma)\sin(2N\varphi')}{\varepsilon_0 \varepsilon_i r_0 \pi \cos(N\pi)} \left[\frac{\min(r_0,\rho')}{\max(r_0,\rho')}\right]^{2n} \left[1 - \frac{1}{2}\delta(N)\right]$$

. (16)

The expressions (16) may be easily evaluated as a geometric series in this configuration, yielding the expression:

-14-

$$\Phi'_i(\rho',\varphi') = \frac{p(\rho'^4 - r_0^4)\sin(\gamma)}{2\pi\varepsilon_0 r_0 \left[\rho'^4 + r_0^4 + 2\rho'^2 r_0^2 \cos(2\varphi')\right]} + \frac{\varepsilon_i p \rho'^2 r_0 \cos(\gamma)\sin(2\varphi')}{\varepsilon_0^2 \pi \left[\rho'^4 + r_0^4 + 2\rho'^2 r_0^2 \cos(2\varphi')\right]}, \qquad (17)$$

which corresponds to the potential distribution of the exciting dipole in Fig. 4 superimposed to a dipole-like singularity placed at $X' = 0$, $Y' = \alpha^2/2R$. This can be seen in the distribution of potential shown in Figure 5 for several orientations of the impressed electric dipole.

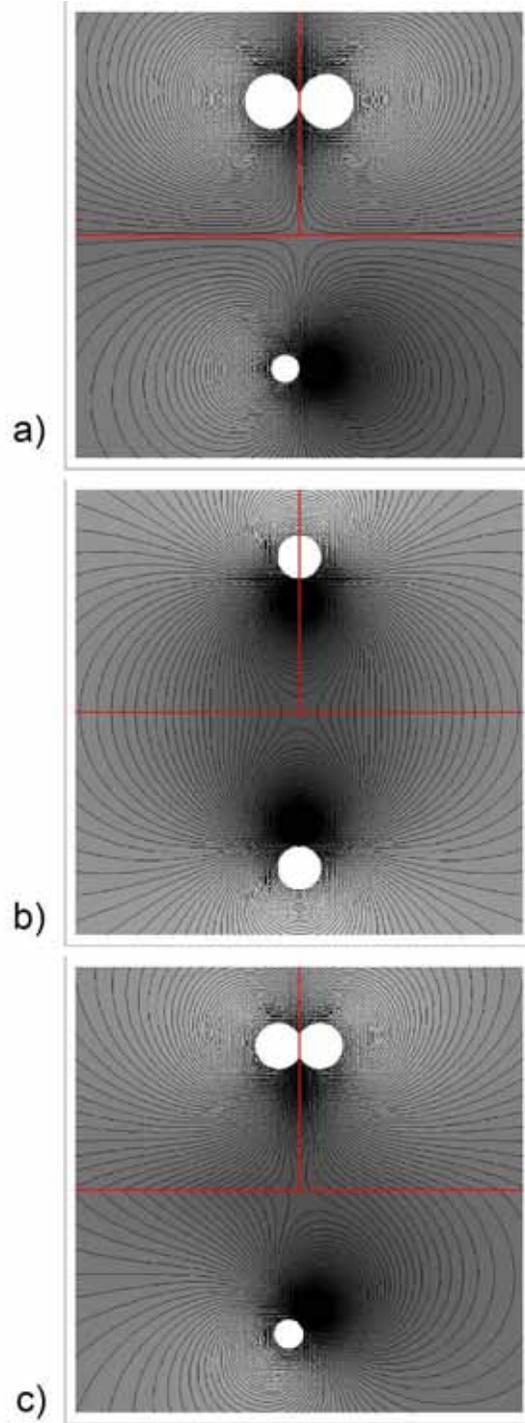

Figure 5 – Potential distributions for a double wedge with permittivities $\varepsilon_1 = -\varepsilon_2 = -2$. The orientation of the exciting dipole is (a) $\gamma = 0$, (b) $\gamma = \pi/2$ and (c) $\gamma = \pi/4$.

We may deduce from Figure 5 some interesting properties of this resonant double wedge configuration. When the impressed dipole is parallel to the external boundary of the double wedge (Fig. 5a), as far as the external field is concerned, the structure behaves as a perfect electric conductor, with the equi-potential line parallel (and therefore the electric field orthogonal) to this boundary. On the other hand, when the impressed dipole is normal (Fig. 5b), as far as the external field is concerned, the structure behaves in a dual way, like a perfect magnetic conductor. In other words, the surface impedance shown by the external interface of the resonant double wedge is strongly affected by the orientation of the impressed dipole and may vary from zero to infinity according to the tangent of the angle $\gamma$. The case of Fig. 5c is an intermediate situation between the two extremes, since the exciting dipole forms an angle of 45 degrees with the boundary. We will see how these phenomena correspond to the series and parallel response of resonant conjoined cylinders as nanocircuit elements in the second part of this paper [17].

## 8. *Inversion of the Kelvin Transform*

The potential distribution for the conjoined cylinder of Fig. 1 may now be evaluated after proper translation and Kelvin inversion of (17), resulting in:

$$\Phi_i(\rho,\varphi) = \Phi_i^{parallel}(\rho,\varphi)\sin(\gamma) + \Phi_i^{series}(\rho,\varphi)\cos(\gamma), \tag{18}$$

where:

$$\begin{cases} \Phi_i^{parallel}(\rho,\varphi) = \dfrac{E_0\left[2R(\rho^2+R^2) - \rho^2 R\cos(2\varphi) - \rho(\rho^2+4R^2)\sin(\varphi)\right]}{\rho^2+R^2-2\rho R\sin(\varphi)} \\ \Phi_i^{series}(\rho,\varphi) = \dfrac{\varepsilon_0 E_0 \rho^2 \cos(\varphi)\left[2R\sin(\varphi)-\rho\right]}{\varepsilon_i\left[\rho^2+R^2-2\rho R\sin(\varphi)\right]} \end{cases}, \tag{19}$$

which is valid in the resonant configuration $\varepsilon_1 = -\varepsilon_2$. The meaning of the superscripts *series* and *parallel* in the previous expressions will be discussed in Part II of this manuscript.

After translating the reference system in such a way that the center of the resonant conjoined half-cylinders coincides with the origin, we get the final simple expressions:

$$\begin{cases} \Phi_i^{parallel}(\rho'', \varphi'') = -E_0 \dfrac{\rho''^2 + R^2}{\rho''} \sin\varphi'' \\ \Phi_i^{series}(\rho'', \varphi'') = -E_0 \dfrac{\varepsilon_0}{\varepsilon_i} \dfrac{\rho''^2 - R^2}{\rho''} \cos\varphi'' \end{cases}, \quad (20)$$

where $(\rho'', \varphi'')$ are the cylindrical coordinates in the new reference system.

Potential distributions for this resonant configuration and physical interpretations and nanocircuit implications of Eq. (20) are provided in [17].

Similar to what was noticed in the case of the rectangular wedge, the split cylinder behaves differently depending on the direction of the incident electric field, which is consistent with their nanocircuit interpretation, as we discuss in Part II. Mathematically this analogy between planar and cylindrical configurations is expected, since the Kelvin transformation is a conformal mapping, which locally preserves the angles, including those between the interfaces and the equipotential lines. The interesting anomalous features of this configuration and their physical interpretation in terms of nanocircuits will be also discussed in detail in Part II.

## *Conclusions*

In this first part of our manuscript, we have presented the closed-form quasi-static analytical solution of the boundary-value problem associated with two conjoined half-cylinders. As we show in [17], this may be interpreted as a series and parallel interconnection of two nanocircuit elements, as a case study of the generalization of our nanocircuit paradigm to take into account coupling and interconnections among nanocircuit elements in a complex nanocircuit. In Part II we will apply the results of this analytical formulation in the nanocircuit paradigm and we will discuss in detail the implications of these results from a physical point of view.